\documentclass[english]{article}
\usepackage{mathptmx}
\usepackage[T1]{fontenc}
\usepackage[latin9]{inputenc}
\usepackage[a4paper]{geometry}
\usepackage{array}
\usepackage{textcomp}
\usepackage{graphicx}
\usepackage{setspace}
\makeatletter

\providecommand{\tabularnewline}{\\}

\newcommand{\lyxaddress}[1]{
\par {\raggedright #1
\vspace{1.4em}
\noindent\par}
}

\makeatother

\usepackage{babel}
\begin{document}

\title{Special electronic structures and quantum conduction of B/P co-doping
carbon nanotubes under electric field using the first principle}

\author{AQing Chen, QingYi Shao%
\thanks{corresponding author(email:qyshao@163.com)%
}, Zhen Li}

\maketitle

\lyxaddress{Laboratory of Quantum Information Technology, School of Physics and
Telecommunication Engineering, South China Normal University, Guangzhou
510006, China}
\begin{abstract}
Boron (B)/phosphorus (P) doped single wall carbon nanotubes (B-PSWNTs)
are studied by using the First-Principle method based on density function
theory (DFT). Mayer bond order, band structure, electrons density
and density of states are calculated. It concludes that the B-PSWNTs
have special band structure which is quite different from BN nanotubes,
and that metallic carbon nanotubes will be converted to semiconductor
due to boron/phosphorus co-doping which breaks the symmetrical structure.
The bonding forms in B-PSWNTs are investigated in detail. Besides,
Mulliken charge population and the quantum conductance are also calculated
to study the quantum transport characteristics of B-PSWNT hetero-junction.
It is found that the position of p-n junction in this hetero-junction
will be changed as the applied electric field increase and it performs
the characteristics of diode.


\end{abstract}

\section{INTRODUCTION}

Since the carbon nanotubes were discovered \cite{iijima_helical_1991},
they have attracted the attention of numerous research groups because
of their outstanding mechanical and electronic properties. A single-wall
carbon nanotube (SWNT) can be described as a graphite sheet rolled
into a cylindrical shape so that the structure is of one dimension
\cite{jose-yacaman_catalytic_1993} with a diameter of about 0.7-10.0
nm. Its electronic structure can be either metallic or semiconducting
depending on its diameter and chirality \cite{dresselhaus_physics_1995}.
At low temperature, a single-wall carbon nanotube is a quantum wire
in which the electrons flow in the wire without being scattered by
scattering centers \cite{charlier_electronic_2007}. They have been
investigated in various fields, especially in carbon nanotubes field-effect
transistors \cite{tans_room-temperature_1998} based on those outstanding
characteristics. What\textquoteright{}s more, chemical doping \cite{lee_conductivity_1997}
is expected to substantially increase the density of free charge carriers
and thereby to enhance the electrical and thermal conductivity. Therefore,
there are a lot of works including theoretical and experimental studies
for doped carbon nanotubes such as boron and nitrogen doped carbon
nanotubes \cite{williams_boron-doped_2007}, \cite{li_effects_2007},
\cite{ishii_resistivity_2008}, \cite{min_unusual_2008}. However,
phosphorus doped carbon nanotubes has a more complex energy band structure
with the presence of two nondispersive P-related bands, one in the
valence band and another in the conduction band \cite{maciel_synthesis_2009}.
In our previous study \cite{chen_effects_2009}, we conclude that
the position of phosphorus in single carbon nanotubes would affect
the electronic structure of SWNTs, which was in agreement with the
results of I. O. Maciel etc \cite{maciel_synthesis_2009}. Although
boron or nitrogen doped carbon nanotubes were synthesized by CVD (chemical
vapor deposition) \cite{ayala_cvd_2008}, \cite{caillard_synthesis_2008},
phosphorus doped carbon nanontubes also was synthesized \cite{maciel_synthesis_2009},
\cite{cruz-silva_heterodoped_2008} and it has the potential for thermoelectric
application. Yong-Ju Kang\textquoteright{}s calculations \cite{kang_electrical_2009}
show that defects such as vacancies severely modify the electronic
structure of carbon nanotubes, resulting in a metal-to-semiconductor
transition. Chemical doping also can convert the metallic SWTN to
semiconductor, which has been discussed by Zhi Xu \cite{xu_converting_2008}
whose study showed that those boron/nitrogen co-doping metallic single
wall carbon nanotubes will be converted to semiconductors. In contrast
to the study of W. L. WANG et al \cite{wang_towards_2007}, our results
demonstrated substitutional P dopants within the nanotube C lattices
is \textit{sp\textsuperscript{\textit{3}}} coordinated rather than
\textit{sp}\textsuperscript{2} coordinated. The constituting B, C
and N elements are homogenously distributed within the SWNTs tube
shells with no appreciable phase-separated B N and C domains, which
have been synthesized by using the Bias-Assisted Hot filament chemical
vapor deposition \cite{wang_direct_2006}. The main difference from
the case of W. L. WANG et al is that B and P elements are implanted
in the C lattices with unhomogeneous distribution in current case.
What's more, the novel feature of such heterojucntion formed by B/P
co-doping is that the position of p-n junction varies as a function
of the electrical field strength. Both Yong-Ju Kang\textquoteright{}s
calculations and Z. Xu studies didn\textquoteright{}t explain the
reason for the transition in detail, but the common characteristic
in their works is that both symmetry of the SWNT structures is broken.
In addition, SWNTs are often synthesized with the mixture of metallic
and semiconducting SWNTs. It is always difficult to separate the semiconductor
SWNTs from metallic ones, which hold back the development of application
of SWNTs. Therefore, the B-P co-doping not only can covert the metallic
to semiconducting SWNTs, but also the B-P doped SWNTs have more complex
electronic structure than B-N doped SWNTs for their special application.
So, it is of a great significance to study the electronic characteristic
of B-P doped carbon nanotubes. So far as we know, however, the characteristics
are not explained in theory in detail. In this paper, we calculate
boron/phosphorus doped carbon nanotubes to explore the reasons for
such a characteristics of metal-to-semiconductor transition deeply
and conclude that symmetric structure broken and \textit{sp}\textsuperscript{3}
hybridization are responsible for metal-to-semiconductor transition.
Another important characteristic of the B-P co-doped SWNT found in
this study is that both B impurity level and one of P impurity levels
are located between the Fermi energy level and the conduction band
but another P impurity level is located between the Fermi energy level
and the valence band, which is quite different from the band structure
of B-N tubes. Besides, Mulliken charge population and quantum conductance
are calculated to characterize quantum transport. It is found that
this hetero-junction displays the structure of n-p-n junction if extra
electrical field is not applied. However, it turns to diode gradually
as the electric field increasing. Besides, the position of p-n junction
shifts to the two layers which are between the two layers of P and
B atoms when the electric field is applied along the tube axis and
increases gradually. Therefore, this study is very useful for solving
the problem of separating the semiconducting carbon nanotubes from
the metallic ones in nano-electronic field. So we consider (9, 0)
and (6, 6) type single carbon nanotubes with 72 and 96 atoms, respectively.
C atoms are substituted with borons and phosphorus in a SWNT. Band
structure, Mayer bond order, density of states Mulliken charge population
and quantum conductance are calculated. Corresponding characteristics
are analyzed to study the characteristics of the hetero-junction further
after a uniform external electric field is applied along the tubes
direction.

\section{MODEL AND THEORETIC APPROACHES}

In this research, a (9, 0) and a (6, 6) type single carbon nanotubes
with 72 and 96 atoms, respectively, are chosen for representing general
metallic tubes to study the characteristic of metallic SWNTs with
different chirality. We substitute four C atoms in carbon framework
with two B atoms two P atoms for (9, 0) type SWNTs, and substitute
two C atoms with a B atom and a P atom for (6, 6) type, respectively,
just as Fig. 1 shows. The chosen tubes are the periodic supercell,
whose band structure is independent of the length of SWNTs representing
the common band structure of doped SWNTs. This consequence can be
proved by comparing Maciel\textquoteright{}s study \cite{maciel_synthesis_2009}
with our previous work \cite{chen_effects_2009}. We use the first
principle based on density functional theory (DFT) which is provided
by DMOL\textsuperscript{3} code \cite{delley_all-electron_1990},
\cite{delley_molecules_2000} available from Accelrys Inc to calculate
the characteristics for showing the feasibility of this hetero-junction.
In order to make the results of calculation reliable and make the
characteristics of such hetero-junction correct, two exchange-correlation
function (the generalized gradient approximation (GGA) for (9, 0)
type and LDA for (6, 6) type) are employed to optimize geometrical
structures and calculate the properties of B-PSWNT with the Perdew-Burke-Ernzerh
\cite{perdew_generalized_1996} and PWC \cite{perdew_accurate_1992}
of correlation gradient correction, respectively, which accompanies
the convergence tolerance of energy of 1.0$\times$10\textsuperscript{-5}Ha
(1Ha=27.2114eV), the maximum force of 0.002Ha/Å, and the maximum displacement
of 0.005Å. The electronic wave functions are expanded in double-numeric
polarized basis set with an orbital cutoff of 5.0Å. The type of treatments
of core electrons is All Electron forms because there are not heavy
atoms in the whole system. The k-points is set 1$\times$1$\times$6
for all structures.

\section{RESULTS}

\subsection{Optimized geometry structure and total energy}

Fig.1 shows the geometry structure of B/P doped SWNTs. It can be seen
that the structure of (6, 6) type is similar to one of (9, 0) type
as Fig. 1 describes. For the geometry structure of (9, 0), two B atoms
and two P atoms substitute for four C atoms located at the corner
of hexagon, respectively. The B atoms are faced to the P atoms directly
with the distance of 2.767Å. P atoms are pulled out of the graphite
sheet. Two reasons can account for that formed structure. One reason
is that the radius of P atom is 
\begin{figure}
\begin{centering}
\includegraphics{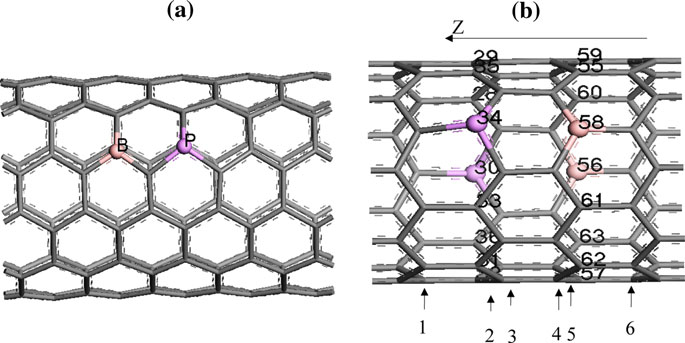}
\par\end{centering}

\caption{Structure of B-P doped SWNT. (a) and (b) are the structure of (6,
6) and (9, 0) type, respectively.}
\end{figure}
larger than that of C atom and another one is that there is more \textit{sp}\textsuperscript{3}
hybridization in C-P chemical bonding. The bond angles of C-P-C are
in the range 99\textasciitilde{}103\textdegree{} and bond lengths
are in the range 1.705 \textasciitilde{}1.798Å. The $\triangle E$
can be used to illustrate the stability of B-PSWNTs under electric
field, which is defined as follows:

\begin{equation}
\triangle E=E^{'}-E
\end{equation}

where $E^{'}$ is the total energy of B-PSWNT under different electric
field and $E$ is the total energy of B-PSWNT without field. Table
I lists the difference between total energy of B-PSWNT without electric
field and that under different electric fields. It can be seen that
the difference of total energies becomes bigger as the electric field
increase. These results suggest that that B-PSWNT may get more stable
under an extra electric field.

\begin{table}
\caption{The difference between total energies of different states}

\begin{centering}
\begin{tabular}{>{\centering}p{4cm}>{\centering}p{4cm}}
\hline 
Type & $\triangle E$( Ha )\tabularnewline
\hline 
B-PSWNT (-0.5V/ Å)  & -0.11\tabularnewline
B-PSWNT (-1V/ Å) & -0.37 \tabularnewline
B-PSWNT (0.50V/ Å) & -0.05 \tabularnewline
 B-PSWNT(1V/ Å)  & -0.24\tabularnewline
\hline 
\end{tabular}
\par\end{centering}

\end{table}

\subsection{energy band structure, density of states and bonding form}

Fig. 2 shows the band structure of intrinsic SWNTs and B-PSWNTs, where
(a) and (b) are band structure of the intrinsic and doped (9, 0) type
SWNTs. A gap about 0.4626 eV appears in Fig. 2 (b). However, Fig.
2 (a) shows that there is no energy gap in intrinsic (9, 0) type SWNT
but an energy level crosses the Fermi energy level. The band structure
of another metallic (6, 6) type SWNTs, just as Fig. 2 (c) shows, is
also calculated in order to illustrate the common phenomenon that
the metallic SWNTs will be converted to semiconductor resulted from
the presence of B and P atoms.There is also a band gap about 0.6 eV
in the band structure of (6, 6) type B-PSWNTs. Therefore, it is concluded
that the metallic SWNT is converted to semiconductor after B/P co-doping,
which agrees well with the results of Zhi Xu, Wengang Lu \cite{xu_converting_2008}.
As far as we know, there is no exact theory to account for such a
characteristic. However, we think the main reasons are that the presence
of impurities destroys the symmetry of geometry structure, which is
viewed as Jahn-Teller effects \cite{jahn_stability_1937}, \cite{janes_metal-ligand_2004},
\cite{Klrner_2001}, \cite{stevenson_observation_2000}, \cite{bearpark_pseudo-jahn-teller_2002}
and that the chemical bonding becomes stronger because of \textit{sp}\textsuperscript{3}
hybridization.

According to some theories \cite{saito_physical_1998}, \cite{datta_quantum_2005}(Satio
et al 2003; Datta 2005), the (9, 0) type SWNT has D$_{9d}$ symmetry
which causes a degeneracy of the energy bands at the boundary of the
Brillouin zone. The energy levels of SWNTs are symmetrically disposed
about E=0 which locate at the six corners of the Brillouin zone. That
is to say that Fermi energy is located at $E=0$. The requirement
of periodic boundary conditions for SWNTs is expressed as follows:

\begin{equation}
\vec{k}\cdot\vec{c}=2\pi\nu
\end{equation}

where $\vec{c}$ is the circumferential vector, which defines a series
of parallel lines, each corresponding to a different integer value
line, giving rise to a set of dispersion relations , one for each
subband $\nu$. The existence of gap depends on whether there exist
subbands passing though the corners of the Brillouin zone or not.
There exist such subbands for metallic SWNTs, but no such subbands
for semiconductors. Therefore, when B atoms and P atoms are injected
into the structure of (9, 0) type SWNT, the symmetry of SWNT will
be broken. So the symmetry of Brillouin zone also is broken. Therefore,
there may be no subbands passing through the corners of the Brillouin
zone and a gap arises as Fig. 2 (b) and (c) show. There is always
an energy gap at the Fermi energy level in one dimension case whenever
the symmetry is broken, which can be concluded from the Jahn-Teller
effects.
\begin{figure}
\begin{centering}
\includegraphics[scale=0.6]{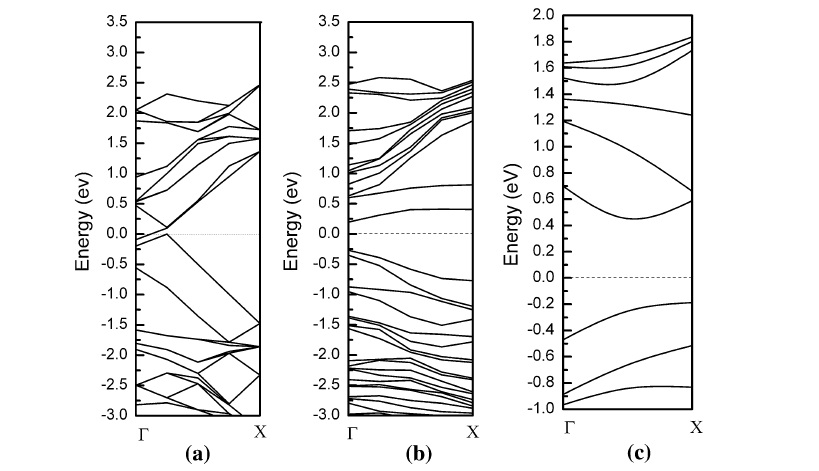}
\par\end{centering}

\caption{(a), (b), and (c) are the band structure of intrinsic (9, 0) type
SWNTs, (9, 0) type B-PSWNTs and (6, 6) B-PSWNTs, respectively. The
Fermi energy is defined at zero.}
\end{figure}

In order to understand the complex electronic structure of B-PSWNTs,
the PDOS of (9, 0) B-PSWNT are calculated just as been shown in Fig.
3. There are two impurity levels due to the presence of P atoms at
\textasciitilde{} 0.4 eV and \textasciitilde{} -0.75 eV and another
impurity level due to the presence of B atoms at \textasciitilde{}
0.7 eV, respectively, by comparing the PDOS of P and B atoms with
the energy band structure in Fig. 3. That is quite different from
the B/N doped SWNTs. The reason for P impurity level dispersion was
explained in our previous study \cite{chen_effects_2009}. Fig. 3
also shows that the values of DOS of C atoms are zero near Fermi energy
level from the LDOS of C atoms. It suggests that there are no energy
levels of C atoms near the Fermi energy level but energy gap comes
out. Another reason considered is that the B/P doped SWNT form a hetero-junction
because of electrons diffusion owing to the difference from electrons
concentrations, which leads to depletion region. It makes the energy
gap get wider.

Mulliken population and Mayer bond order \cite{mulliken_electronic_1955},
\cite{mayer_bond_1986} were calculated to understand the bonding
form. The result data are listed in Tab II. The form Am is used to
indicate the atoms, where subscript m is the ordinal of atoms. The
Mayer total valences of atoms from the second layer to the fifth layer
are listed in the Table II. The second to fifth layer is corresponding
to 2st\textasciitilde{}5th layer of the structure (see Fig. 1), respectively.
\begin{figure}
\begin{centering}
\includegraphics[scale=0.6]{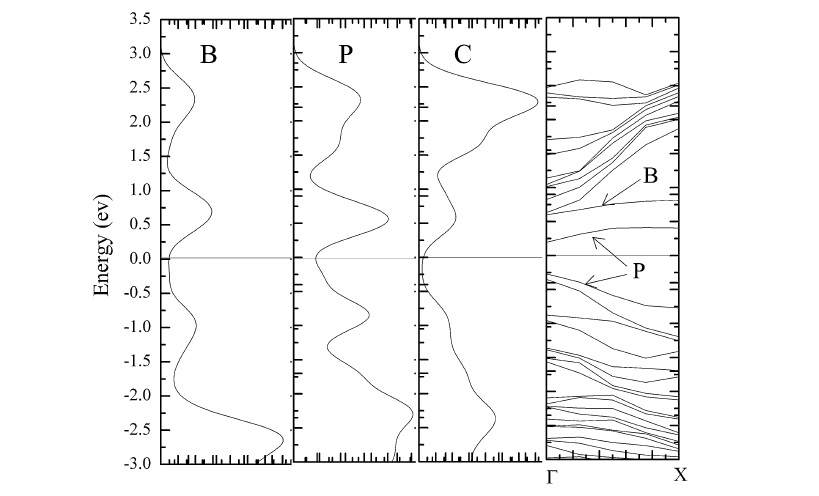}
\par\end{centering}

\caption{Energy band structure of B-PSWNTs and LDOS of B, P and C atom, respectively.
The Fermi energy is defined at zero.}
\end{figure}
 Both the valences of P$_{30}$ and of P$_{34}$ are 3.2, while the
valence of B$_{56}$ is 3.1 as the same as that of B$_{58}$. The
total valence indicates how many single bonds are associated with
the atom \cite{mayer_bond_1986}. Therefore, both P atom and B atom
form about 3 bonds. The values of bond orders of P atoms forming bonding
with three C atoms are 0.99, 0.99 and 0.90, while those of B atoms
forming bonding with three C atoms are 1.03, 1.04 and 0.94. Being
different with above cases, those of C atoms forming bonding each
other are in the range 1.11\textasciitilde{}1.35. The \textit{sp}\textsuperscript{3}
hybridization mainly forms a single bond. Therefore, there is more
\textit{sp}\textsuperscript{3} hybridization in P atom than that
in B atoms, which is the same as that in C atoms. Therefore, it can
be concluded that there is more \textit{sp}\textsuperscript{3} hybridization
in B/P doped SWNT than that in intrinsic SWNT, which may be another
important cause of presence of an energy gap. The Mayer total valence
of each C atom in the 2nd, 3rd, 4th and 5th layer is closed to four
indicating each C atom with four bonds. 
\begin{table}
\caption{Mayer total valence of C, B, and P atom in layer 2-5 respectively}

\medskip{}

\centering{}%
\begin{tabular}{cccc}
\hline 
No.2 layer Mayer & No.3 layer Mayer  & No.4 layer Mayer & No.5 layer Mayer\tabularnewline
 total valence & total valence & total valence & total valence\tabularnewline
\hline 
C 28 3.5  & C 1 3.6  & C 46 3.6  & C 55 3.5 \tabularnewline
C 29 3.5  & C 2 3.7  & C 47 3.6  & B 56 3.1 \tabularnewline
P 30 3.2  & C 3 3.7  & C 48 3.7  & C 57 3.5 \tabularnewline
C 31 3.5  & C 4 3.7  & C 49 3.6  & B 58 3.1 \tabularnewline
C 32 3.5 & C 5 3.6  & C 50 3.6  & C 59 3.6 \tabularnewline
C 33 3.5  & C 6 3.6  & C 51 3.6  & C 60 3.5 \tabularnewline
P 34 3.2  & C 7 3.6  & C 52 3.6  & C 61 3.6 \tabularnewline
C 35 3.5  & C 8 3.7  & C 53 3.7  & C 62 3.6 \tabularnewline
C 36 3.5 & C 9 3.6 & C 54 3.6 & C 63 3.6\tabularnewline
\hline 
\end{tabular}
\end{table}

The deformation density is calculated to analyze the bonding forms
further. Fig. 4 shows the deformation density of intrinsic (9, 0)
type SWNT and that of B/P doped SWNT. Deformation density of the layers
B and P atoms locate at are Fig. 4 (a) and (b), respectively. Fig.
4 (c) displays the deformation density of the layer which C atoms
locate at in intrinsic SWNT. It is observed that the deformation density
is homogeneous in Fig. 4 (c). However, the deformation density is
homogeneous inside of annulus but not well-proportioned outside for
B-PSWNTs as been shown in Fig. 4 (a). There are fewer electrons in
the red area called bonding area, which gets wider after boron and
phosphorous are injected. It indicates that there is more \textit{sp}\textsuperscript{3}
hybridization in that area. It also can be seen that the deformation
density is not homogeneous inside and outside of the annulus in Fig.
4 (b). It suggests that there is more \textit{sp}\textsuperscript{3}
hybridization in the bonding formed by P atoms and C atoms than that
formed by B atom and C atom, which agrees well with the analysis of
Mayer bond order. The bonds of B-PSWNT become stronger in comparison
to the similar bonds of intrinsic SWNT. According to the distribution
of deformation density in Fig. 4 (a) and (b), there are lots of delocalized
orbitals. Therefore, the electron transfer will happen in B/P doped
SWNT. The space charge region is produced when the electron transfer
happens. 
\begin{figure}
\begin{centering}
\includegraphics[scale=0.55]{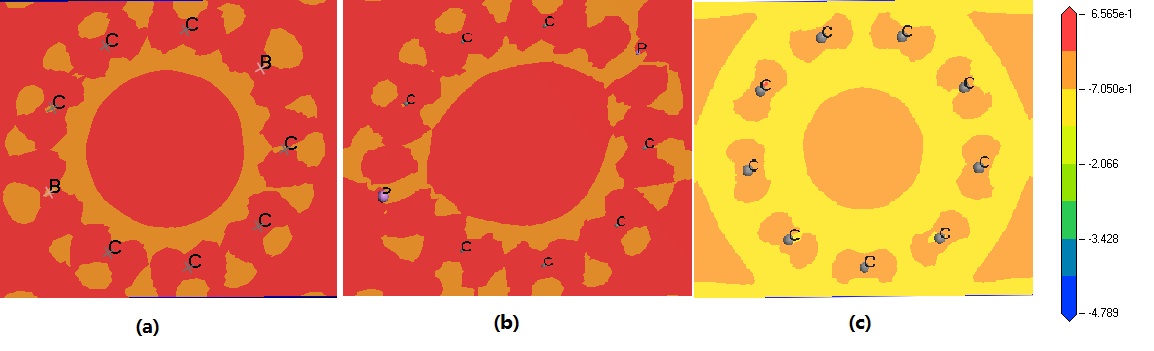}
\par\end{centering}

\caption{(a), (b) and (c) are the deformation density in the layers which B,
P and C atom locate at respectively. A scale locates at the right
of figure.}

\end{figure}

\section{Quantum conductance}

Quantum conductance is calculated to study the quantum transport of
this hetero-junction. Carbon nanotubes are the quasi one dimensional
nano-wire. The electrons are transported in the form of ballistic
transports when the length of nano-wire is shorter than the mean free
path of an electron under low temperature. Ballistic transport consists
of single electron conduction with no phase and momentum relaxation.
When chemical dopants are introduced into the C lattice, the impurity-induced
backscattering becomes severely strong \cite{biel_anomalous_2009},
yielding a very small mean free path so that electrons are strongly
backscattered on the impurity level. However, quasiballistic behavior
is restored under the proper condition according to work R. Avriller\textquoteright{}s
work \cite{avriller_chemical_2006} which mainly studied the effect
of tube length and the magnetic field on the quantum conductance of
CNTs. Doping by physisorption or chemical substitutions made great
impact on the quantum transport of CNTs. Ch. Adessi's study \cite{adessi_reduced_2006}
which demonstrated that doping by physisorption will decrease charge-carrier
mobilities much less than chemical substitution introducing strong
backscattering at the resonance energies of the\textquotedblleft{}quasibound\textquotedblright{}
states associated with the defect potential well. Three type transport
regimes including quasi-ballistic, intermediate and localized regime
have been studied by R. Avriller et al \cite{avriller_low-dimensional_2007}.
The case of (9, 0) considered in this study is quite different from
the case R. Avriller group studied. Comparing R. Avriller\textquoteright{}s
work in which characteristic transport length scales like the elastic
mean free path and the localization length were studied, the length
of this tube is about 7.032 Å which is great shorter than the mean
free path or the elastic mean free path (about 17.8 nm in this case).
Therefore, our study focus on the ballistic transport of B-P doped
SWNTs. The feature of quantum transport of the p-n junction region
is main study in this section. The quantum conductance $G$ can be
attained by Landauer formula \cite{datta_quantum_2005}, \cite{van_wees_quantized_1988}
which demonstrates that quantum conductance $G$ is determined by
the number of the energy level across the Fermi energy level. The
Landauer formula is written as follows:

\begin{equation}
G=G_{0}M
\end{equation}

\begin{equation}
G_{0}=q^{2}/h=38.7\mu S=\left(25.8k\Omega\right)^{-1}
\end{equation}

where $M$ is the number of energy level across the Fermi energy level.
Since the number of sub-energy band equal to the integral value of
the density of states, the distribution of $G$ under various electric
fields as a function of energy range can be described as Fig. 5. 
\begin{figure}
\begin{centering}
\includegraphics{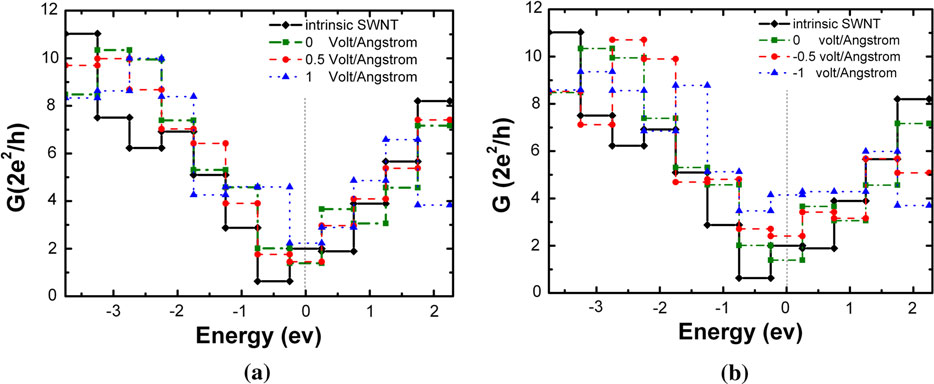}
\par\end{centering}

\caption{(a) and (b) display the quantum conductance of (9, 0) case, as figure
1 (b) defined, under positive and negative electric filed, respectively.
The quantum conductances of both intrinsic and B-P doped SWNT are
described in each figure. Ef = 0 (vertical dotted line) is given.}
\end{figure}
A uniform extra electric field is applied along the direction of the
tubes and the positive direction is defined as Z direction (see Fig.
1) to calculate the characteristic of electron transport. The range
of electric field strength varies from -1.0 to1.0 V/Å. The quantum
conductance is shown in Fig. 5. We define that the electric field
is positive along the Z direction (see Fig. 1). Therefore, Fig. 5
(a) and (b) show the quantum conductance of (9, 0) case under the
positive and negative electric field, respectively. Fig. 5 (a) shows
that the quantum conductance of intrinsic (9, 0) type SWNT is 2$G_{0}$
and that of B/P doped SWNT is about 1.3$G_{0}$ when electric field
strength is 0V/Å. Substitutional doping in CNTs will induce backscattering
efficiency which reduced charge mobility and will strongly impact
on the on-current capability of CNTs \cite{adessi_reduced_2006}.
But the main reason for the decrease of the conductance is effects
of the build-in field formed by B/P co-doping. The blow phenomenon
can well interpret such reason. When electric field strength increases
up to 0.5V/Å, there are not few changes of $G$. As electric field
increases to 1V/Å, the quantum conductance $G$ is up to 2$G_{0}$.
Therefore, figure 5 (a) suggests that the hetero-junction of B/P SWNTs
may be broken down under the electric field with 1 V/Å according to
varied value of quantum conductance as a function of electrical field.
This can be explained by the Tunneling Effect because the junction
described in Fig. 6 becomes narrow under the electric field with 1
V/Å. Just as Fig. 5 (b) described, when the electric field strength
is -0.5 V/Å, the quantum conductance $G$ is closed to 3$G_{0}$.
What\textquoteright{}s more, the quantum conductance $G$ will increase
to 5$G_{0}$ under the electric field with -1 V/Å. The potential barrier
weakens as the electrical field is increased positively so that the
quantum conductance increases. While the electrical field becomes
negative the carries can be transported under the electrical field.
It is therefore concluded that the B/P doped SWNT behaves like p-n
junction.
\begin{figure}
\begin{centering}
\includegraphics{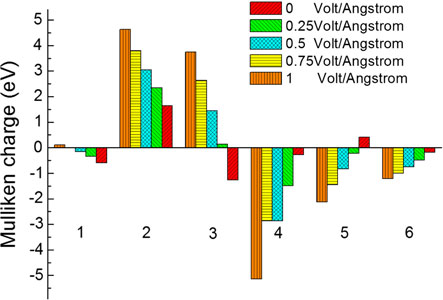}
\par\end{centering}

\caption{The Mulliken charge population of B-PSWNTs. The charge of each layer
is shown under different electric field, where the layer number is
defined in Fig. 1.}

\end{figure}

The electronic distribution in the B-PSWNTs is determined by Mulliken
charge population as the Fig. 6 shows. The electronegative of C atom
is larger than that of B atom resulting in the fact that the fifth
layer B atoms locate in are positive and that the nearest neighbor
layers are in reverse. However, the second layer P atoms locate in
is positive because P atoms give electronics to the C atoms to take
part in \textit{sp}\textsuperscript{3} hybridization, which is quiet
different from the fifth layer. It can be seem that the p-n junction
main locates between the second and the third layer without applying
extra electric field while the position of p-n junction shift to the
interface between the third and the fourth layer gradually as electric
field increase. Therefore, the p-n junction has the tendency to be
formed in the layers between the layers in which P and B atoms locate.
It is important to note that the length of p-n junction get longer
when the electric field is 0.25 V/ Å. Since the charge of the second
layer is nearly zero, which may leads to decrease of the conduction.
But the length will become shorter with the electric field increase
resulting in the increase of conduction. The above results agree well
with what the Fig. 5 demonstrates.

\section{CONCLUSION}

By calculating the characteristics of B-P doped (9, 0) and (6, 6)
type SWNT using the first principle based on DFT, it concludes that
the metallic SWNT will convert to semiconductor because of the B/P
co-doping. The main reasons are that the symmetry is broken and that
there are more sp3 hybridizations in B/P doped SWNT than those in
intrinsic SWNT. It is also seen that the band structure of B-PSWNTs
is special owing to the special band structure of P doped SWNTs. Lastly,
Mulliken charge population and the quantum conductance of that hetero-junction
are calculated to study the electrons transport. The study of quantum
conductance suggests that this hetero-junction has the characteristics
of p-n junction. Finally, it found that the position of p-n junction
of B-PSWNTs will be changed under the electric field by the Mulliken
charge population analysis, which will be significance for SWNTs application
in nano-device in the future. 

\bibliographystyle{unsrt}
\bibliography{cite}

\end{document}